**Comparing Retrieval Strategies to Capture Interdisciplinary Scientific Research: A Bibliometric Evaluation of the Integration of Neuroscience and Computer Science**

Malena Méndez Isla[*1,2] https://orcid.org/0000-0002-9748-2709,
Agustín Mauro[3] https://orcid.org/0000-0002-6052-3757, &
Diego Kozlowski[1] https://orcid.org/0000-0002-5396-3471

[1] Université de Montréal, École de bibliothéconomie et des sciences de l'information, Montréal, Canada.
[2] Consejo Nacional de Investigaciones Científicas y Técnicas (CONICET), Centro de Estudios Urbanos y Regionales, Buenos Aires, Argentina.
[3] Consejo Nacional de Investigaciones Científicas y Técnicas (CONICET), Universidad Nacional de Córdoba, Instituto de Humanidades, Córdoba, Argentina.
[*] corresponding author: malena.mendez.isla@umontreal.ca

**ABSTRACT**

Interdisciplinary scientific research is increasingly important in knowledge production, funding policies, and academic discussions on scholarly communication. While many studies focus on interdisciplinary corpora defined a priori—usually through keyword-based searches within assumed interdisciplinary domains—few explore interdisciplinarity as an emergent intersection between two distinct fields. Thus, methodological proposals for building databases at the intersection of two fields of knowledge are scarce. The goal of this article is to develop and compare different strategies for defining an interdisciplinary corpus between two bodies of knowledge. As a case study, we focus on the intersection between neuroscience and computer science. To this end, we develop and compare four retrieval strategies, two of them based on keywords and two based on citation and reference patterns. Our results show that the reference-based strategy provides better retrieval, pseudo-recall, and F1. While we focus on comparing strategies for the study of the intersection between the fields of neuroscience and computer science, this methodological reflection is applicable to a wide range of interdisciplinary domains.

## 1. Introduction

Interdisciplinary scientific research (IDR) is increasingly present in scholarly communication and collaboration (Vienni Baptista et al., 2020; Klein, 2021) as well as in science policies (Schmidt, 2021; Woods et al., 2024). IDR is often viewed as a means to enhance scientific progress and its societal impact; as such, it is frequently associated with funding requirements (Omodei et al., 2017) from organizations such as the National Science Foundation (NSF), the UK Research Councils, the National Natural Science Foundation of China, or the Swedish Research Council, among others (Wang, 2016). Nevertheless, IDR is also seen as a contested concept (Klein, 2021), and the construction of empirical evidence regarding its impact is a persistent challenge (Marres & De Rijcke, 2020; Rafols & Meyer, 2010).

IDR is a type of research that integrates data, techniques, tools, concepts, or theories from two or more bodies of knowledge to enhance fundamental understanding or to address problems beyond a single discipline's scope (National Academies, 2005). This research aims to generate results that offer a new understanding of a phenomenon (Wagner et al., 2011). While multidisciplinarity refers to the sum or the addition of knowledge from different disciplines without an integration (Klein, 1990), interdisciplinarity is greater than the sum of knowledge from different fields. Interdisciplinarity is, therefore, a process of knowledge integration, which may include social and cognitive aspects (Klein, 2008). Transdisciplinarity refers to a process where researchers work together to develop theoretical frameworks that transcend scientific boundaries, focusing on how to solve societal problems by integrating knowledge from various scientific and social bodies of knowledge; thus, transdisciplinarity integrates non-scientists in the research process (Vienni Baptista et al., 2020).

Bibliometric methods allow researchers to study IDR, focusing on scientific outputs. The increase in interdisciplinary research has influenced a growing literature (Wagner et al., 2011), including different ways to conceptualize and measure IDR. There is no consensus on categorization methods for measuring interdisciplinarity, and due to its multidimensional nature, no single method of analysis can reveal all of its aspects (Chang & Huang, 2012). Previous literature can be classified into two methodological approaches (Morillo et al., 2001; Sugimoto & Larivière, 2018): 1. co-authorship and 2. citation- or reference-based analysis. Co-authorship refers to those methods that have authors as their unit of analysis and focus on collaborations between authors from different fields (Porter et al., 2008; Schummer, 2004). This is associated with small-scale works due to the difficulty of assigning each author to a single discipline, and therefore not much work has been done from this perspective (Porter et al., 2007; Sugimoto & Larivière, 2018). The citation- or reference-based approach has papers as units of analysis and is quite common (Garfield, Malin & Small, 1978; Klein, 2008), as its operationalization is more straightforward (Wagner et al., 2011).

Although in this paper we analyze the concept of IDR to consider the conjunction of two specific disciplines, to a large extent the literature in bibliometrics on the subject concentrates on the development of ways to measure the degree of interdisciplinarity of a paper. Focusing on this problem, another way to classify bibliometric literature on IDR is to distinguish between 1. Top-down and 2. Bottom-up perspectives (Rafols & Meyer, 2010). Top-down perspectives rely on discipline classifications provided by databases, while bottom-up approaches define an emergent structure based on co-words, co-citations, or bibliographic coupling (Wagner et al., 2011). The top-down or classification-based perspectives calculate the degree of interdisciplinarity of articles using references and citations (Porter et al., 2007). These works usually use ISI subject categories and the Rao-Stirling index (Rao, 1982; Stirling, 2007), which provides a framework for evaluating IDR by considering three key attributes of diversity: Variety, which refers to the number of distinct disciplines cited in an article's references. Balance, which refers to how even the distribution of citations is across various disciplines, and disparity or similarity, which assesses the extent of the difference between the cited disciplines (Sugimoto & Larivière, 2018). Leydesdorff (2019) proposes a new indicator of diversity by calculating variety, balance (using the Gini coefficient), and disparity separately and combining them a posteriori.

In the bottom-up, or spatial perspectives, papers are classified into factors based on multivariate analyses of similarity measures. For example, by creating a network of journals through co-citation and using betweenness centrality as an indicator of IDR (Leydesdorff, 2007). Rafols & Meyer (2010) combine top-down and bottom-up approaches, using diversity measures with large-scale disciplinary categories and network measures based on publication similarities, applying bibliographic coupling.

While we start from conceptualizing the IDR between two fields, most of the previous literature in bibliometrics starts from a specific research area, which they define a priori as IDR. Those specific areas of study are then retrieved via journals and keywords. This conceptualization is applied to research areas such as medical education (Maggio et al., 2023), behavioral economics (Truc, 2022), hate speech (Ramírez-García et al., 2022), or robotics in surgery (Shen et al., 2019).

In particular, Costas & Bordons (2008) tested different ways of defining marine science as an IDR, and compared retrieval approaches based on journals and the combination of journals and keywords. Archambault et al. (2009) studied the field of energy research and development using a search strategy that combined the use of journals, keywords, references, and citations. The following two works study precision and recall in citation-based strategies using systematic reviews as a gold standard. Bascur et al. (2023) evaluated citation-based clusters and found their performance highly variable, highlighting that clusters work best when recall is prioritized over precision (ratio 2-8) and point out that citation-based clusters can complement query-based search by retrieving additional relevant documents. Sjögårde and Ahlgren (2024) find that co-citation outperforms direct citation and bibliographic coupling and suggest combining citation-based and textual methods to enhance retrieval performance.

To our knowledge, only a few studies focus on systematically developing a strategy to define an interdisciplinary corpus between two bodies of knowledge. That is to say, not a specific interdisciplinary topic that combines two fields, such as brain-computer interface, but all the possible subjects that could possibly represent the emergence of a combined discipline. Delineating a scientific domain is a preliminary task in most bibliometric studies, for which considerable bibliometric research has been done and for which there are roughly three strategies: ready-made classifications of science, classical information-retrieval searches, and mapping and clustering techniques (Zitt et al., 2019). But those are targeted at finding a specific scientific domain only; thus, it is a disciplinary delineation.

Given that delineating the corpus is one of the basic tasks in a bibliometric study, delineating an interdisciplinary corpus improves the possibilities for studies on IDR. Karunan et al. (2017) represent one of the closest efforts, using two approaches: cross-citations and overlapping papers. Thus, they identify two types of interactions: the ones that occur through boundary papers and the ones that manifest themselves through cross-citations. However, Karunan et al. (2017) does not test for precision or recall and neither test the efficacy of thematic filters nor the role of references in delineating IDR.

This gap underscores the need for a systematic approach to building databases at the intersection of disciplines. In this article, we differentiate our work from previous studies by testing and comparing a keyword-based approach with a citation- and reference-based one. Although bibliometric studies in recent years have focused on the latent content of texts, the semantic approach has not been applied to IDR at the intersection of two bodies of knowledge. We do that on the one hand by considering keywords in the delimitation of an interdisciplinary corpus and, on the other hand, in the topical analysis of the documents using the BERTopic model (Grootendorst, 2022).

As a case study for the different methods, we will focus on the intersection between computer science and neuroscience. Recently, Simard et al. (2023) showed that over the past 30 years, the number of brain-related papers has grown faster than science in general, with China leading the way. Simard et al. (2023) also point out that this Chinese preponderance is partly due to the inclusion of articles from computer science in the neuroscience sample. This posed a question about the current intersection of these fields of knowledge. Neuroscience and computer science have historically exchanged theories, concepts, techniques, instruments, methods, and data. For example, McCulloch & Pitts (1943) proposed modeling neurons as logic gates, opening up a whole field of exploration for computer science inspired by neuroscience. The interaction between the two disciplines extends beyond computational neuroscience, which focuses on modeling brain processes, and neuroinformatics, which applies computational techniques to neuroscientific data. There are many subdisciplines where the interaction and exchange between neuroscience and computer science are key. Therefore, a more comprehensive approach is needed to build a database that captures all the kinds of interactions between the two disciplines.

The goal of this article is to develop and compare different strategies for defining the intersection between two bodies of knowledge. Focusing on the case study of the fields of neuroscience and computer science, we developed four strategies for delimiting an interdisciplinary corpus and compared the potential and limitations of each. The strategies focus on keywords (co-occurrence and cross-keywords) and citations (cross-citations and cross-references). To compare these four strategies, we developed the following research questions:

RQ1: How do the strategies vary in terms of retrieval, overlap, and time evolution?
RQ2: To what extent are disciplines, journals, and research topics represented in each strategy?
RQ3: How do the strategies differ in terms of precision and pseudo-recall?

The main contribution of this paper lies in discussing and comparing the relative performance of the four strategies proposed. The article is structured as follows: In the Methods section, we describe each strategy in conceptual and methodological terms. In the Results section, we present and compare the performance of the strategies. Finally, in the Discussion section, we evaluate the advantages and limitations of each strategy.

## 2. Data & Methods

### Data

For this research, we used data from the Web of Science (WoS), accessed by the Observatoire des Sciences et des Technologies (OST), between 1991 and 2023, given the availability of keywords.

To define the fields, we used Hamilton's (2003) classification for the definition of computer science, which is based on the framework established by the NSF. For computer science, we used the predefined speciality "Computers". For neuroscience, that strategy is not possible, as there is not a "neuroscience" category. The closest classification is the specialty "Neurology & Neurosurgery", which does not include all neuroscience publications. To start with two comparable corpora, we first defined neuroscience by selecting a set of journals that structure the field in the same way as Hamilton (2003) defines computer science. To do this, we searched for journals that contain the wildcards "brain*" or "neuro*" in their titles or that belong to the specialty "Neurology & Neurosurgery". We retrieved a dataset of 564 journals, which was then curated by excluding three unrelated journals. The final definition of the field of neuroscience consists of 561 journals. Based on the computer science and neuroscience journals list, we extracted the corresponding articles.

To create the sets of keywords we selected the top 1,000 **WoS authors' keywords** from each discipline, ranked by absolute frequency. These sets include both high-retrieval and less frequently used terms, with the least frequent keyword appearing in 316 documents. For a detailed discussion of alternative keyword ranking methods and the rationale behind our final choice, see Appendix 1.

From the top 1,000 we manually curated a list of 100 authors' keywords for each discipline using a terminological disambiguation criterion, prioritizing semantic clarity. Our goal was to retain terms that were unambiguous. We excluded terms commonly used in both fields but with different meanings. For example, the term *neural network* was excluded as it is polysemic. In neuroscience, it refers to the biological networks of interconnected neurons in the brain, whereas in computer science, it denotes artificial architectures used in machine learning, inspired by—but fundamentally different from—biological systems. Although the term is the same, its conceptual grounding, applications, and disciplinary contexts differ significantly. Including such terms could have led to misclassifications of papers as interdisciplinary, when in fact they remain within the boundaries of a single discipline. In other words, ambiguous terms could lead to false positives in the identification of interdisciplinary research. For a discussion of potential residual ambiguity and its implications for interdisciplinary measurement, see the Limitations section. Although the process was guided by clear disambiguation criteria, subjective judgment was unavoidable. To mitigate this, selections were made in the framework of a meeting of authors to ensure consistency and reduce individual bias.

As a result, we obtained a list of 100 keywords for neuroscience and another for computer science (see Appendix 2), where the least frequent keyword appeared in 460 documents. This approach captures both mainstream interdisciplinary fields and more recent and evolving areas of integration. While we prioritized frequency to ensure robustness, our selection process did not exclusively favor long-established or mainstream topics. The curated list of keywords includes several emerging or less consolidated topics that have gained visibility in recent years. For example, keywords such as *fog computing, edge computing, brain modeling, and neuromodulation* reflect novel or expanding areas of research at the intersection of computer science and neuroscience. These terms may not yet dominate the overall publication landscape, but their frequency signals a growing presence in the interdisciplinary discourse.

## Methods

We propose two retrieval approaches: the keyword-based approach and the citation- and reference-based approach. The keyword-based (or semantic) approach includes two strategies: Strategy 1, shared keywords, which retrieves papers containing both neuroscience and computer science keywords, and Strategy 2, cross-keywords, which retrieves papers from neuroscience with computer science keywords and vice versa. The citation-based (or relational) approach consists of Strategy 3, which focuses on cross-citations, and Strategy 4, which targets cross-references. In the following lines, we describe the rationale behind each strategy:

**Strategy 1 (S1): Shared keywords.** This strategy retrieves papers that contain keywords from both neuroscience and computer science without using the disciplinary corpora of articles. S1, by definition, returns papers from any discipline. This makes it a strategy that combines a bottom-up approach (since there is no predefined disciplinary classification) with a top-down approach, given that the keywords originate from the neuroscience and computer science definitions. Most research on measuring IDR focuses on references and citations, making S1 less directly aligned with previous studies than the other three strategies. However, it relates to content-based approaches, which have gained prominence in bibliometrics in recent years. Additionally, S1 shares similarities with studies that predefine an interdisciplinary area based on the presence of specific keywords in titles (Maggio et al., 2023; Ramírez-García et al., 2022; 2019). In addition, S1 allows the retrieval of papers that other strategies fail to bring because it is not bounded by the two disciplines, which also carries the risk of retrieving inaccurate articles, which we evaluate in this study.

**Strategy 2 (S2): Cross-keywords.** This strategy retrieves papers from the neuroscience corpus that use keywords from the computer science corpus and vice versa. In this case, we use two inputs previously described: the corpora of neuroscience and computer science and the keyword lists. S2 combines semantic integration at the boundaries of disciplines, focusing on how fields of knowledge interact through shared terminology. S2 contributes to the aim of this work by allowing us to study the linguistic aspect of IDR together with the disciplinary aspect without relying on references or citations. This approach aligns with recent trends in bibliometrics that emphasize content-based methods to complement traditional citation-based studies.

**Strategy 3 (S3): Cross-citations.** This strategy retrieves articles from the neuroscience corpus that contain citations from the computer science corpus and vice versa. S3 contributes to the aim of this research by focusing on how one field's work is recognized through its appropriation by another field. It is, in part, aligned with emerging bottom-up perspectives on measuring IDR that are using citations (Wagner et al., 2011). In this strategy, citations serve as the indicator of IDR. Thus, it highlights one aspect of the phenomenon while excluding papers without citations. We introduced a minimal filtering condition in Strategy 3, requiring that papers have at least two citations overall, as it maximizes F1. Figure 1 shows the evolution of precision, the relative recall with respect to no thresholding, F1 (using relative recall), and the total number of cases retained by the strategy for a random sample of 200 cases labelled for this fine-tuning. Our results show that excluding papers with only one citation improves the precision without affecting recall. We also tested the threshold for a minimum proportion of citations (and of references for the strategy 4) from the opposite field, but the results showed that these thresholds did not improve precision, while they negatively affected recall (see Appendix 3).

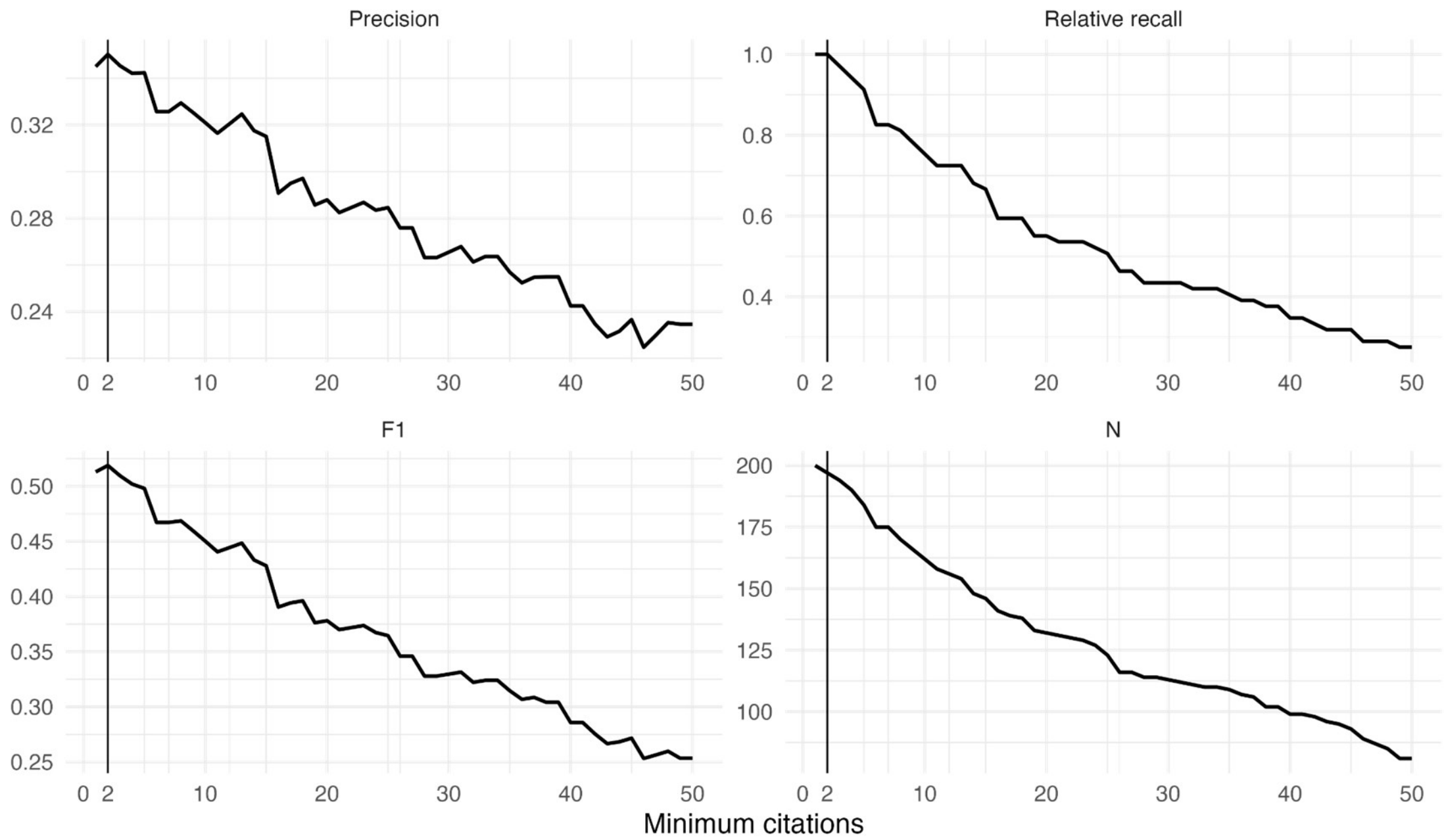

**Figure 1.** Effect of total citation thresholds on Strategy 3's performance metrics. Relative recall is based on the no-threshold scenario.

**Strategy 4 (S4): Cross-references.** This strategy retrieves articles from each disciplinary corpus that contain references to the other disciplinary corpus. S4 contributes to the objective of this work, accounting for the *use* of literature from the other field. References are a common method to measure IDR, focusing on the degree of interdisciplinarity. We tested thresholds based on the proportion of references to the opposite field, but these did not improve precision and reduced recall, so we did not apply any threshold (see Appendix 3).

While S2, S3, and S4 have a top-down direction, given the categorization of the NSF for computer science and our definition in neuroscience, S1 is bottom-up since it allows articles from any discipline to be retrieved. However, it retains a top-down component since it starts with keywords derived from the fields' definitions.

## Research topics

To provide a thematic representation of the interdisciplinary corpus (Kozlowski et al., 2021), we adopted an approach for research topic analysis using pre-trained NLP embeddings, specifically the BERTopic model, which represents the current state of the art. BERTopic (Grootendorst, 2022) is a topic model that extracts a coherent topic representation through the development of a class-based variant of TF-IDF. BERTopic is a multilingual, semi-supervised method for topic modeling that generates document embeddings with pre-trained transformer-based language models, clusters these embeddings, and finally, generates topic representations with the class-based TF-IDF procedure (Grootendorst, 2022). By leveraging all the articles from the entire corpus, we construct the thematic representation through the application of BERTopic over abstracts and titles. We applied BERTopic to a minimum cluster of 500 papers, resulting in the identification of 78 research topics.

## Precision and pseudo-recall

To determine the precision of each strategy, we started reviewing random samples of 500 papers per strategy. We read titles and abstracts, discussed best practices, and created guidelines for deciding inclusion or exclusion (see Appendix 4). Our main criterion was that the integration of knowledge from both fields was meaningful for the article's research aims and clearly stated in the title or abstract. We identified three main types of relations.

1. Articles based on neuroscience that used tools from computer science, such as the use of machine learning for disease detection in neuroimaging studies. In this category, routine applications of established computational methods were excluded, and only those cases were classified as positive where there was a clear objective to develop or advance neuro-computational approaches representing novel contributions.
2. Articles based in computer science that applied knowledge from neuroscience, for instance, to develop brain-inspired algorithms.
3. Articles that were not clearly based in one field but still exhibited a meaningful integration of knowledge, such as those on brain-computer interfaces.

With this criteria, we analyzed new random samples of 500 papers from each strategy and manually labeled them to estimate precision and pseudo-recall.

To understand the false positives from the strategies, we provide selected examples organized by type of misclassification and by strategy. We identified three main types of false positives:

1. **Neuroscience-only papers without computer science components** (e.g., S1: "A model of dopamine modulated cortical activation"; S2: "Vitamin D deficiency in multiple sclerosis: Should testing and treatment be based on racial background?").
2. **Computer science-only papers without neuroscience relevance** (e.g., S3 and S4: "An Effective and Novel Neural Network Ensemble for Shift Pattern Detection"; S4: "Image identification and labeling using hybrid transformation and neural network").

3. **Papers applying computer science to health domains beyond neuroscience** (e.g., S4: "A Novel Deep Learning Network and Its Application for Pulmonary Nodule Segmentation").

The strategies show distinct misclassification patterns: S1 and S2 tend to include neuroscience papers lacking a computer science component, and those applying computer science beyond neuroscience; S3 often retrieves computer science-only papers; S4 retrieves both non-neuroscience health applications and computer science papers lacking neuroscience components.

**False negatives**

During the manual validation, we also identified false negatives. As we do not have access to a gold standard corpus, we operationalized false negatives, which are retrieved by at least one strategy, but missed by another. There are three main types of false negatives:

1. **Computer science applications to neuroscience** (e.g. S4: "White Matter Connectome Edge Density in Children with Autism Spectrum Disorders: Potential Imaging Biomarkers Using Machine-Learning Models")
2. **Neuro-inspired computer sciences** (e.g. S1 and S2: "A Unified Biologically-Inspired Prediction Framework for Classification of Movement-Related Potentials Based on a Logistic Regression Model")
3. **Hybrid papers such as those on brain-computer interfaces** (e.g. S2: "Wireless Brain-Computer Interface for Electric Wheelchairs with EEG and Eye-Blinking Signals")

While S3 and S4 have false negatives primarily on computer science applications to neuroscience, S1 fails to retrieve neuro-inspired computer science papers, and S2 misses hybrid papers, such as those involving brain-computer interfaces.

To compute the recall of each strategy, we would need to know the actual population of the neuro-computer science paper, which is unknown. Instead, we computed a pseudo-recall by combining all the manually checked positive cases and calculating the ratio of positive cases for each strategy to the total pool of positive cases. We performed bootstrap resampling to calculate the confidence interval of precision and recall using 10,000 replications.

## 3. Results

### 3.1. Retrieval, overlap and time evolution

Figure 2, presents the number of articles retrieved by each strategy, their overlap and their evolution over time.

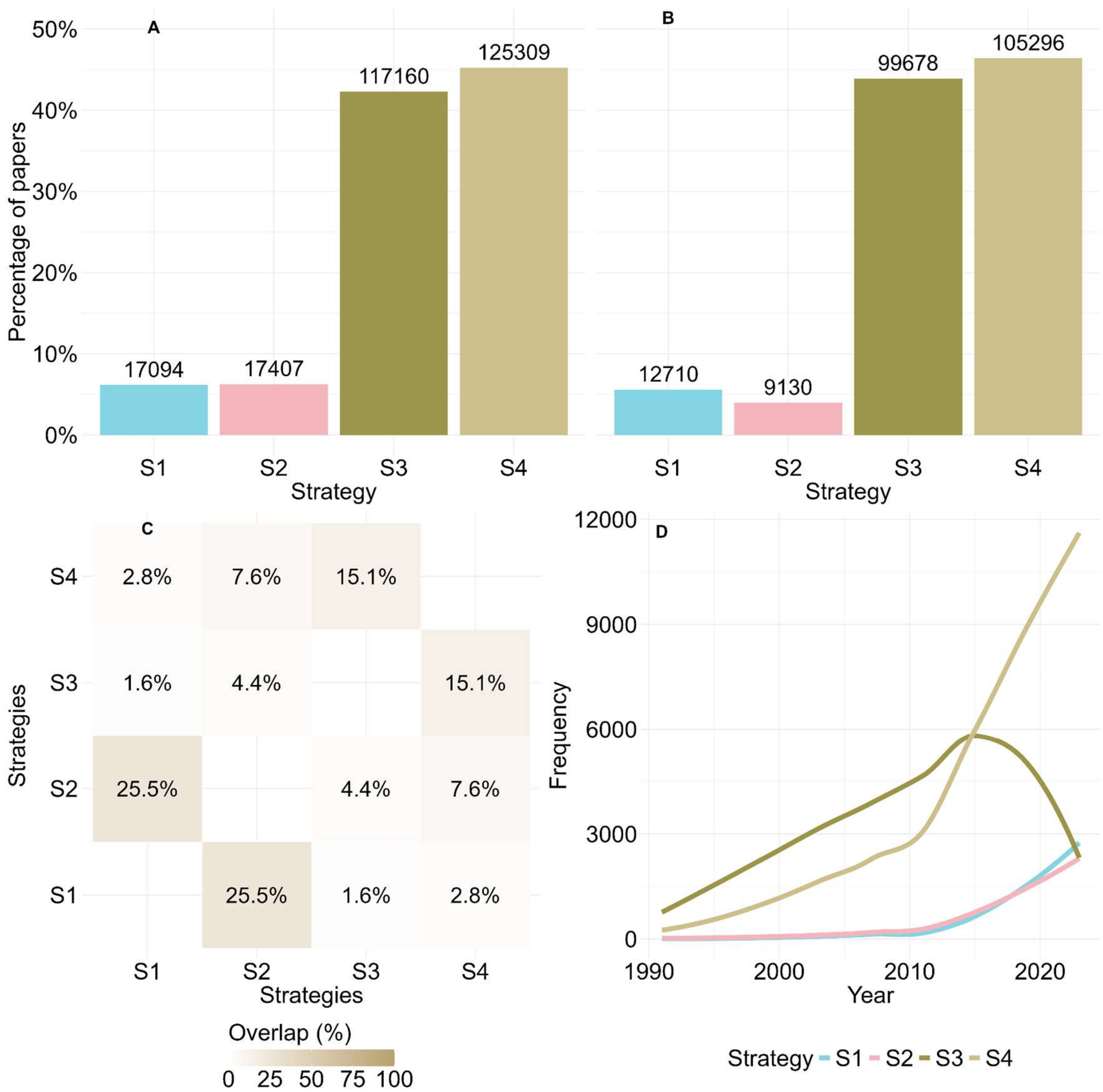

**Figure 2: Retrieval power by strategies, overlap, and number of articles per year (1991–2023).** Panel A shows the absolute number of articles retrieved, while Panel B presents fractionalized counting for articles retrieved by multiple strategies. Panel C depicts the overlap between strategies. Overlap (%) values are based on the Jaccard index, calculated as the size of the intersection divided by the size of the union of documents retrieved by each pair of strategies. Panel D illustrates the temporal evolution of the number of articles retrieved per year for each strategy. The line shows the relation between the variables with a locally weighted scatterplot smoothing fit (Lowess).

Panel A (absolute counting) shows that citation- and reference- based strategies have the greatest retrieval, with S4 leading the way (125,309), followed by S3 (117,160), S2 (17,407), and S1 (17,094).

Although S1 searches across all disciplines in Web of Science, it requires articles to match at least two keywords from a relatively constrained list. In contrast, S2 operates within the narrower neuroscience and computer science corpora but retrieves articles based on a single keyword match, which results in broader inclusion. This difference in matching criteria may account for the higher number of articles retrieved by S2, despite its more limited disciplinary scope.

S3 and S4 go beyond lexical matching, relying instead on citation patterns to identify papers that draw meaningfully from both fields. This enables them to capture more implicit or conceptually integrated forms of interdisciplinarity, particularly in areas where terminology is not standardized or where one field adopts methods or perspectives from another without using its vocabulary.

Panel B shows the fractional counting of retrieved documents for each strategy, where documents retrieved by more than one strategy are counted as a fraction. This figure shows the higher retrieval volumes of citation-based strategies (S3 and S4), which reflect their broader scope: these strategies identify interdisciplinary connections based on citation behavior, even when no explicit lexical overlap is present. In contrast, keyword-based strategies (S1 and S2) return smaller but more selective sets. S1 brings a different set of articles, even if it does not return the largest collection. This can be attributed to its unique construction method, which does not rely on a disciplinary base like the other strategies, enabling it to include novel articles exclusive to this strategy.

Panel C shows the proportion of shared documents across strategies. First, we see that the strategies with a shared approach (semantic or citation-based) show a larger overlap. Semantic based strategies (S1-S2) have a 25% overlap, while citation-based strategies have a 15% overlap. It is striking how low the overlap between semantic- and citation-based strategies is, between 1.6% and 7.6%. This highlights how different strategies capture very different corpus.

Panel D shows the number of documents retrieved by each strategy over time. There has been a tendency for all strategies to increase over time, indicating that the intersection between neuroscience and computer science is growing, regardless of the strategy used. S4 shows a rising tendency since 2010, similar to the tendency of S1-S2.

S3 shows a slower growth and a decrease of that same period. This is a reflection of the time it takes citations to accumulate, which are the signal used by strategy S3. The minimum number of citations' threshold introduced to S3 to improve its performance (see Figure 1) only worsen that temporal bias. Even if S3 retrieves one order of tenfold more documents than S1-S2 (Panel B), this temporal bias is problematic as it creates an artifact where the overall tendency observed in this interdisciplinary field is misrepresented, as it seems to be decreasing over time, when it actually increases.

### 3.2. Disciplines

Figure 3 shows the distribution by field per strategy. There, we can see that computer science ("Engineering and Technology") is more involved in strategy 4, while neuroscience ("Clinical Medicine") is more involved in strategies 1, 2 and 3. S1 includes a smaller number of papers from computer science and a greater variety of disciplines.

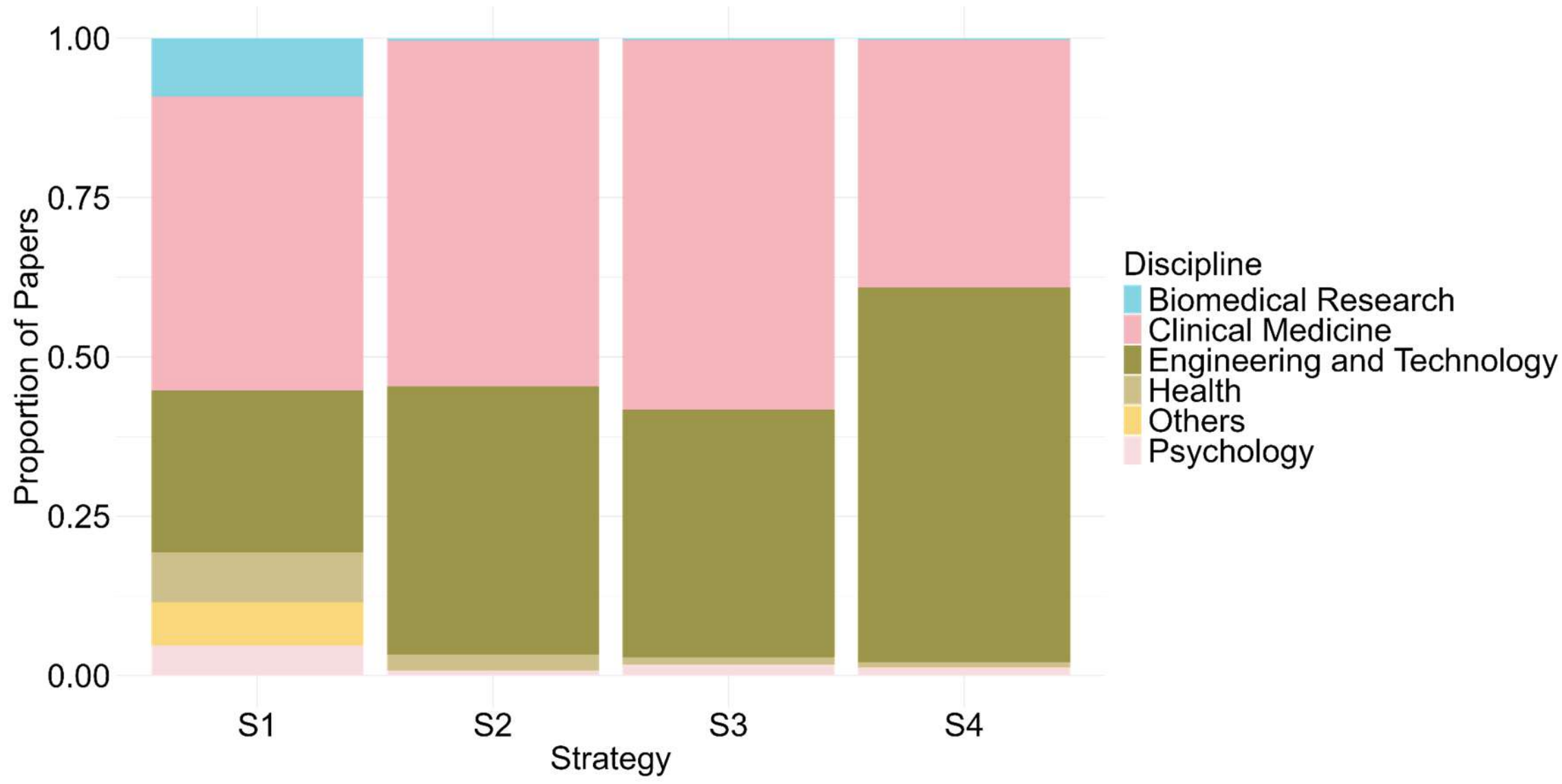

**Figure 3: Distribution of disciplines across strategies**

S1, which is not discipline-based, brings together 14 disciplines due to its construction method. Among them, “Clinical medicine” (46%), “Engineering and technology” (25%), “Biomedical research” (9%), “Health” (8%), and “Psychology” (5%) stand out. The category 'Others,' only present in the case of S1, includes Physics, Professional Fields, and Mathematics, among others. S2 has 54% of its articles on “Clinical medicine - Neurology and Neurosurgery” and 42% on “Engineering and technology - Computers.”. S3 includes 39% % of articles from “Engineering and Technology—Computers” and 58% from “Clinical Medicine—Neurology and Neurosurgery.” S4 retrieves 59% of its articles from “Engineering and technology - Computers” and 39% from “Clinical Medicine—Neurology and Neurosurgery.” Overall, we observe that S1 offers a greater disciplinarity variety and S2 is the most balanced between the two main fields. S3 is also relatively balanced, with a slight emphasis on neuroscience whereas S4 shows a majority of articles from computer science.

I

### 3.3. Research Topic Analysis

We use all articles from all corpora to build a thematic representation with the BERTopic model (see Methods section). Then we compute the distribution of topics across strategies, as shown in Figure 4, to understand the ability of each strategy to capture different research topics. Figure 4 presents a heatmap showing the distribution of topics across strategies, calculated as the percentage of articles in each strategy that belong to each topic.

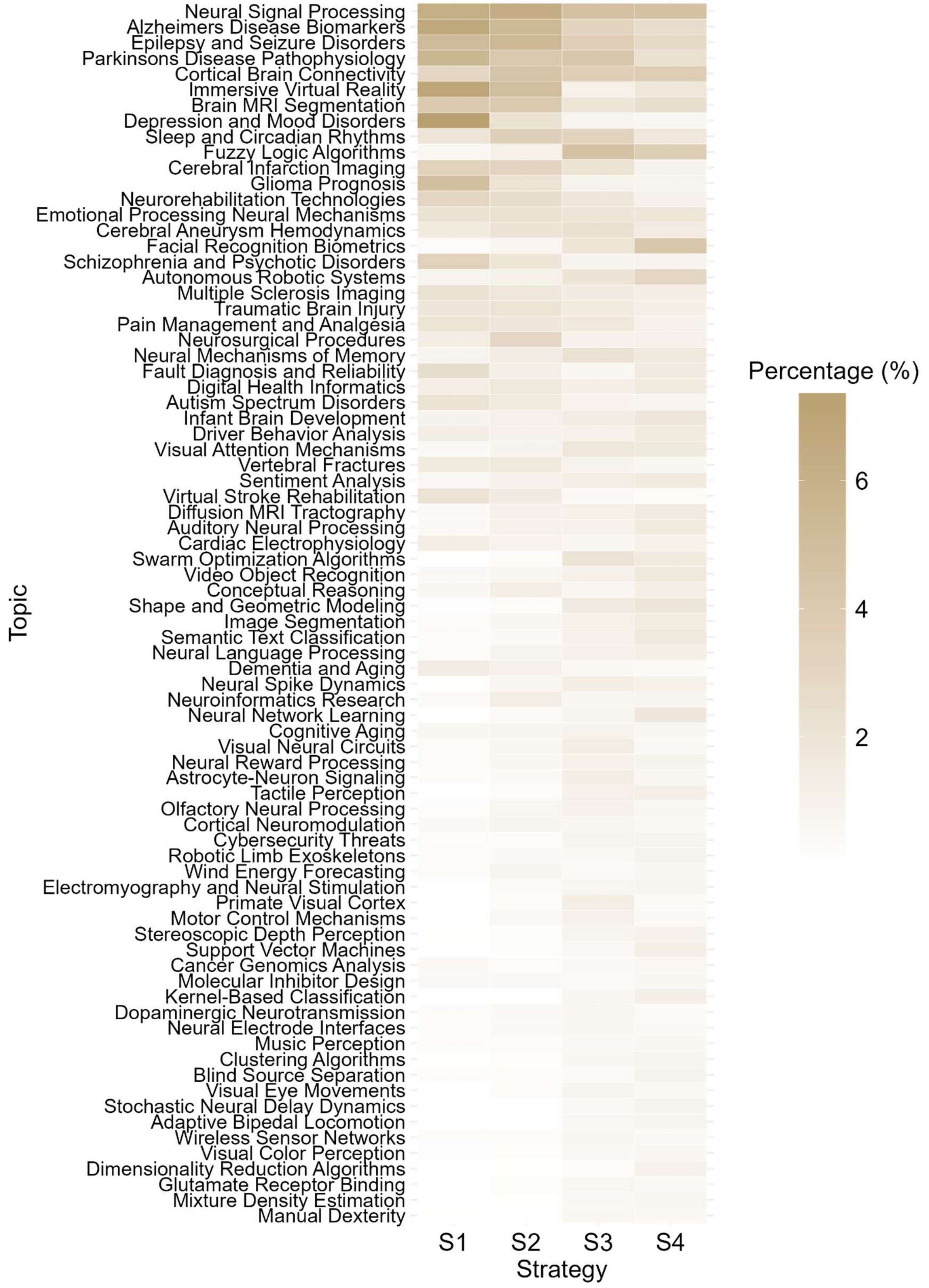

**Figure 4. Distribution of topics across strategies, calculated as the percentage of articles in each strategy that belong to each topic.**

Figure 4 shows that disease-related topics are covered to markedly different extents across strategies. By aggregating the topics and their number of publications we see that S1 stands out with the highest share of papers focused on diseases (36%), followed by S2 (18%) and S3 (15%), while S4 shows the lowest presence of disease-related content (7%). This indicates a much stronger emphasis on disease-focused research in S1 (that is a keyword-based strategy defined by the most frequent terms) compared to the other strategies. For instance, papers in S1 frequently address topics such as “Depression and Mood Disorders”, “Alzheimer’s Disease Biomarkers”, “Parkinson’s Disease Pathophysiology”, “Epilepsy and Seizure Disorders”, “Schizophrenia and Psychotic Disorders”, “Glioma Prognosis”, and “Cerebral Infarction Imaging”. S2 and S3 also include some of these topics, though to a lesser extent. In contrast, S4 focuses predominantly on technology-based topics, with less representation of disease-related ones. Notably, “Depression and Mood Disorders”—the most frequent disease topic in S1, accounting for 7% of its papers—represents only 2.2% of S2, 0.8% of S3, and 0.7% of S4 papers.

The topic “Neural Signal Processing” ranks first in S2 (6%), S3 (5%) and S4 (5%). This topic bridges fundamental questions about brain function with advanced computational methods, including machine learning techniques for analyzing and modeling neural data. It includes the decoding of brain activity to understand cognitive, motor, or sensory processes, the design of algorithms and models to interpret neural patterns under normal or pathological conditions, the development of brain-computer interfaces (BCI), neural prosthetics, or neurofeedback systems, and the diagnosis and monitoring of neurological disorders, such as epilepsy, Parkinson’s disease, or sleep disorders.

While we acknowledge that defining keyword-based strategies using the top 100 most frequent terms may introduce a bias toward mainstream and well-established interdisciplinary areas, our topic-level analysis helps to qualify this effect. S1, for example, shows a strong focus on widely studied disease-related topics, suggesting a higher sensitivity to the core of interdisciplinary production. Moreover, S3—despite being citation-based—also appears biased toward more established topics, as evidenced by its 15% share of disease-focused papers and its more limited coverage of recent themes. In contrast, S4 retrieves a more diverse set of topics, including emerging and technology-oriented areas. Overall, S1 emerges as the most oriented toward dominant areas of research, S2 and S3 appear less constrained by mainstream trends, while S4 stands out as the most open to novelty and peripheral areas.

We also explored a possible alignment between topics and the categories identified during manual validation. Specifically, we manually assigned each topic to one of the five categories defined in the Methods section: (1) neuroscience-based articles that employed tools from computer science (*CS-applied-to-NS*), (2) computer science-based articles inspired by neuroscience (*neuro-inspired*), (3) *hybrid* articles that did not clearly originate in one field but demonstrated meaningful integration of both domains, (4) *NS-only* articles with no identifiable computer science component, and (5) *CS-only* articles with no clear relevance to neuroscience. As the classification in this case is made over topics instead of articles, we cannot directly infer that every article that falls under categories *NS-only* or *CS-only* is a false positive, as there could be articles within those topics that fit our guidelines for the manual validation (see section 3.3 for the manually validated articles). Nevertheless, this broader classification allows us to compare the distribution over these categories for each strategy.

Based on this classification, we calculated the number of papers in each strategy that fall into each topic category. The results of this categorization are presented in the table below.

**Table 1.** Number of papers per strategy associated with each topic category. Categories were assigned at the topic level through manual validation.

| Strategy | CS-applied-to-NS | Hybrid | Neuro-inspired | NS-only | CS-only |
|---|---|---|---|---|---|
| S1 | 27% | 36% | 2% | 30% | 4% |
| S2 | 25% | 42% | 3% | 25% | 6% |
| S3 | 15% | 50% | 3% | 19% | 13% |
| S4 | 15% | 50% | 5% | 13% | 17% |

These results suggest that computational neuro-inspired topics are generally scarce. It is also plausible that neuroscience articles tend to adopt more computer science-related keywords—particularly in strategies like S1 and S2, which appear more sensitive to retrieving articles where computational techniques are applied from within neuroscience. These articles may be less likely to be captured by citation-based strategies such as S3 and S4. Conversely, hybrid articles—arguably more representative of core interdisciplinary integration, where disciplinary boundaries are blurred—seem to be better retrieved by S3 and S4 than by S1 and S2. While these interpretations remain speculative, they offer a useful perspective for understanding how different retrieval strategies may favor certain types of interdisciplinary configurations over others.

### 3.4. Precision and pseudo-recall

The performance of each strategy was assessed by manually validating random samples of 500 papers to ensure meaningful knowledge integration between both fields (see Appendix 4). Confidence intervals were estimated using bootstrap resampling with 10,000 replications. We also computed the pseudo-recall over all the positive cases manually evaluated across the four strategies (see Data & Methods). Table 2 shows the precision, pseudo-recall, and the confidence intervals (CI) for both metrics.

**Table 2: Comparison of strategies according to precision and pseudo-recall.**

| **Strategy** | **Precision & CI**<br>**N 500** | **Pseudo-recall & CI**<br>**N 500** | **F1** |
|---|---|---|---|
| S1 | 60%<br>(56%-63%) | 15%<br>(13%-17%) | 23% |
| S2 | **61%**<br>**(57% - 66%)** | 20%<br>(17%-23%) | 30% |
| S3 | 36%<br>(32%-40%) | 60%<br>(56%-64%) | 45% |
| S4 | 45%<br>(41%-49%) | **75%**<br>**(71%-78%)** | **56%** |

The manual validation shows that S2 has the best precision (61%), closely followed by S1 (60 %). S4 has a precision of 45%, and S3 shows the lowest performance with a precision of 36%. S4 has the best pseudo-recall performance (75%) followed by S3 (60%), S2 (20%) and S1 (15%). F1 shows that S4 (56%) is the most balanced strategy followed by S3 (45%), S2 (30%) and S1 (23%).

Our precision results align with—but are slightly lower than—those obtained by Costas & Bordons (2008) for retrieving interdisciplinary marine science papers: 69% precision and 75% recall using a hybrid journal- and keyword-based strategy and 96% precision with 30% recall for a journal-based strategy. This discrepancy is expected given the greater complexity of the task: Retrieving interdisciplinary papers spanning two distinct fields is inherently more challenging than delimiting a single interdisciplinary area selected with a thematic criterion. Similarly, Archambault (2009) achieved 87% recall and high manually verified precision when studying the energy R&D field using a comprehensive strategy combining journals, keywords, citations, and references.

Our results for citation-based strategies echo those of Bascur et al. (2023), who found that the performance of citation-based clustering varies significantly with parameter choices. For example, using $\beta = 4$, they reported a median precision of approximately 20% and recall around 40%, suggesting that such methods may favor recall over precision. These findings also align with those of Sjögårde and Ahlgren (2024), who reported co-citation-based approaches achieving a precision of 15% and a recall of 65%.

**Limitations**

This study has several limitations. First, the relatively low values of precision and pseudo-recall across all strategies reflect the intrinsic complexity of identifying IDR. These metrics are, in part, a consequence of the strict criteria applied during manual validation, where we excluded marginally relevant documents. For example, we excluded neuroscience articles that employed computational techniques to measure specific phenomena when the use of such methods was routine and instrumental, rather than constitutive of the article's novelty or research objective. While our performance values fall within the range reported in the literature (e.g., Costas & Bordons, 2008; Archambault et al., 2009; Bascur et al., 2023), we recognize that the task remains inherently difficult, and that our definitions and criteria may contribute to the modest retrieval metrics.

Second, some strategies may have excluded relevant IDR publications from high-impact journals. Strategy 1 accessed the full WoS database and could, in principle, retrieve papers from multidisciplinary journals such as *Nature*, *Science*, and *PNAS*. However, the

keyword list used to build S1 was also based on journal classifications, which may have limited the ability of S1 to fully capture papers from these sources. Strategies 2 to 4 were further restricted to specific disciplinary categories; thus, they may have omitted high-impact IDR papers published in multidisciplinary journals not classified within those categories. Although some compensation was achieved by retrieving articles from specialized journals within the same editorial portfolios (e.g., *Nature Neuroscience, Science Robotics*), the exclusion of top multidisciplinary journals remains a constraint. We acknowledge that this limitation highlights the need for future work to capture IDR across *Nature*, *Science*, and *PNAS*, among others.

Third, we acknowledge limitations in our keyword selection process. To ensure robustness, we excluded keywords that appeared in fewer than 460 documents, which may have led to the omission of novel or emerging topics that have not yet gained widespread visibility. Additionally, while we manually curated the top 1000 most frequent WoS authors' keywords per field to minimize polysemy, some semantic ambiguity may still persist. For instance, although we excluded terms with widely divergent meanings across fields (e.g., "neural networks"), we retained certain keywords like "artificial neural networks" to preserve their relevance within important interdisciplinary areas. This residual polysemy could affect the accuracy of Strategies 1 and 2. We therefore acknowledge that some interdisciplinary noise may remain in the data, despite our efforts to reduce it.

Fourth, a limitation inherent to citation-based strategies like S3 is their inability to capture recent publications that have not yet accumulated citations—this issue would persist even without any filtering. On top of this, the citation threshold applied in S3 (requiring at least two total citations) exacerbates the underrepresentation of recent records. For instance, in 2021 there were 5,906 total records, of which 5,418 met the threshold, leaving 488 (~8%) excluded; in 2022, 1,949 total with 1,434 included, excluding 515 (~26%); and in 2023, 19 total with 12 included, excluding 7 (~37%). This reflects the natural lag in cumulative citations for more recent publications. As a result, the temporal evolution analysis shows a downward trend since 2021.

Fifth, our study is limited by the use of the WoS database, which entails language, geographic, and disciplinary restrictions (Aguado López & Vargas Arbeláez, 2016) Some contributions in computer science are published as patents or conference proceedings rather than papers, which are not considered in WoS. All the strategies have a bias against non-English-speaking authors, given the predominance of English-language content in WoS.

Finally, our evaluation lacked a gold-standard dataset for IDR. This remains a central challenge in the field. To address it, we constructed a manually validated benchmark using strict and transparent selection criteria that emphasize genuine integration across disciplines. While our efforts contribute to the foundation of such a benchmark, further work is needed to develop standardized datasets and consensus definitions for IDR evaluation.

In sum, these limitations underscore the need for cautious interpretation of our findings. At the same time, they reflect broader challenges in operationalizing and measuring IDR, and point to key areas for future methodological improvement.

## 4. Discussion

This study compared four strategies for IDR data retrieval at the intersection of neuroscience and computer science. While strategies S1 (shared keywords) and S2 (cross-keywords) are semantic-based, strategies S3 (cross-citations) and S4 (cross-references) rely on social networks. Keyword-based approaches demonstrated higher precision while

citation- and reference-based strategies demonstrated higher recall. The F1 metric suggests that S4 is the most balanced strategy. However, the highest values of precision (61% for S2) and pseudo-recall (75% for S4) may appear relatively low. From our perspective, these low performance metrics reflect the intrinsic complexity of the task and the strict criteria applied during manual validation (see Appendix 4). For instance, neuroscience papers that use routine computational techniques to solve neurosciences disciplinary questions were not considered interdisciplinary in our evaluation. Nevertheless, the levels of precision and pseudo-recall obtained in this work are within the range reported in the cited research, which oscillates between 15% and 69% for precision, and 40% and 87% for recall (Archambault et al., 2009; Bascur et al., 2023; Costas & Bordons, 2008; Sjögårde & Ahlgren, 2024).

This work represents a first rigorous attempt to propose systematic strategies for addressing the problem of retrieving IDR, applied to the case of the intersection between neuroscience and computer science. We acknowledge the limitations of the current strategies, including decisions related to keyword selection. However, we believe that by making the design choices explicit and providing a benchmark dataset, this study contributes by providing a starting point for future refinement and validation. The implications of our findings should indeed be interpreted with caution, and we see this work as an invitation for further empirical and methodological advances in the retrieval of IDR.

The advantages of S1 (shared keywords) are that it allows for retrieving articles from any discipline, aligned with emerging bottom-up perspectives. S1 primarily represents topics focused on diseases, which may be related to the fact that this strategy is based on the most frequent keywords. . In terms of overlap, it overlaps with S2 and has the lowest overlap with S3 and S4, as S1 itself does not start from predefined disciplines, which serve as the basis for the other strategies. Other advantages of S1 are that it may partly address the challenge of journal classification and that it offers a bottom-up approach.

The gains of S2 (cross-keywords) include having the best performance in precision.S2 has the best overlap with other strategies, outperforming S1 in this regard. Another advantage of S2 is that it focuses on content, similar to much of the bibliometric work conducted in recent years. S2 accounts for 18% of papers addressing disease-related research topics. The precision performance of the keyword-based perspective, when defined by the most frequent terms, is an emerging finding of this exploratory research.

The strengths of S3 (cross-citations) lie in its alignment with emergent citation-based perspectives (Rafols & Meyer, 2010). S3 includes 15% of disease-related topics and seems to be able to retrieve *hybrid* articles that did not clearly originate in one field but demonstrated meaningful integration of both domains. One limitation of S3 is that it could lead to false negatives by missing out on highly interdisciplinary papers, which, as Larivière and Gingras (2010) point out, tend to receive fewer citations. Another limitation for S3 is its lack of inclusion of recent publications. These potential limitations appear to be an issue with S3, as it has the lowest precision compared to the other strategies. Additionally, compared to S4, S3 is less open to novel and emerging topics, reflecting a stronger bias toward established areas of research.

The advantage of S4 (cross-references) is that it aligns with the reference-based approach to measuring IDR, which has been widely used in bibliometrics (Stirling, 2007; Leydesdorff et al., 2019). Additionally, it includes non-cited articles, unlike S3, which is citation-based and excludes articles with few or no citations. S4, like S3, accounts for the social aspect of IDR, but without the limitations of citation-based strategy. S4 predominantly reflects technology-oriented topics and tends to retrieve articles that, rather than originating from a single discipline, exhibit substantial integration across domains. S4 has the greater retrieval, pseudo-recall, and F1.

The topic-level analysis shows potential for future research lines as topics can potentially show a more direct or indirect relation with the interdisciplinary discussions. Although our topic classification was not intended for retrieval purposes, the alignment observed between certain topics and the manually validated categories suggests that topic modelling could refine or complement existing strategies, particularly it could be used as a second order filtering to exclude topics that are only from Neuroscience or Computer Science. This needs further exploration as the topics are inferred and there might be relevant documents within what seems to be a false positive topic.

The lack of manual evaluation of performance for the construction of interdisciplinary corpora between two fields highlights the need for this type of research, as it critically reflects on a crucial step for bibliometric analysis. Our study highlights methodological considerations for retrieving interdisciplinary research, particularly at the neuroscience-computer science intersection. While reference-based strategy outperformed citation- and keyword-based approaches (S3 and S4), their precision and recall values underscore the importance of information retrieval evaluation in bibliometric studies of interdisciplinary research.

## References

Aguado López, E., & Vargas Arbeláez, E. J. (2016). Reapropiación del conocimiento y descolonización: El acceso abierto como proceso de acción política del sur. *Revista Colombiana de Sociología*, *39*(2), 69–88.

Archambault, É., Caruso, J., Côté, G., & Larivière, V. (2009). *Bibliometric Analysis of Leading Countries in Energy Research*.

Bascur, J. P., Verberne, S., Van Eck, N. J., & Waltman, L. (2023). Academic information retrieval using citation clusters: In-depth evaluation based on systematic reviews. *Scientometrics*, *128*(5), 2895–2921. https://doi.org/10.1007/s11192-023-04681-x

Beigel, F. (2013). Centros y periferias en la circulación internacional del conocimiento. *Nueva Sociedad, 245, 110-123.* https://doi.org/10.3989/redc.2008.v31.i2.428

Chang, Y., & Huang, M. (2012). A study of the evolution of interdisciplinarity in library and information science: Using three bibliometric methods. *Journal of the American Society for Information Science and Technology*, *63*(1), 22–33. https://doi.org/10.1002/asi.21649

Costas, R., & Bordons, M. (2008). Development of a thematic filter for the bibliometric delimitation on interdisciplinary area: The case of Marine Science. *Revista española de Documentación Científica*, *31*(2), 261–272. https://doi.org/10.3989/redc.2008.v31.i2.428

Garfield, E., Malin, M., & Small, H. (1978). Chapter 8 Citations Data as Science Indicators (pp. 179–208). In *Toward a metric of science. The advent of science indicators. New York: John Wiley.*

Grootendorst, M. (2022). *BERTopic: Neural topic modeling with a class-based TF-IDF procedure* (No. arXiv:2203.05794). arXiv. http://arxiv.org/abs/2203.05794

Hamilton, K. S. (2003). *Subfield and level classification of journals (CHI Research No. 2012-R).*

Karunan, K., Lathabai, H. H., & Prabhakaran, T. (2017). Discovering interdisciplinary interactions between two research fields using citation networks. *Scientometrics*,

*113*(1), 335–367. https://doi.org/10.1007/s11192-017-2481-0

Klein, J. T. (1990). *Interdisciplinarity: History, theory, and practice.* Detroit: Wayne State University Press.

Klein, J. T. (2008). Evaluation of Interdisciplinary and Transdisciplinary Research. *American Journal of Preventive Medicine*, *35*(2), S116–S123. https://doi.org/10.1016/j.amepre.2008.05.010

Kozlowski, D., Dusdal, J., Pang, J., & Zilian, A. (2021). Semantic and relational spaces in science of science: Deep learning models for article vectorisation. *Scientometrics*, *126*(7), 5881–5910. https://doi.org/10.1007/s11192-021-03984-1

Leydesdorff, L. (2007). Betweenness centrality as an indicator of the interdisciplinarity of scientific journals. *Journal of the American Society for Information Science and Technology*, *58*(9), 1303–1319. https://doi.org/10.1002/asi.20614

Leydesdorff, L., Wagner, C. S., & Bornmann, L. (2019). Diversity measurement: Steps towards the measurement of interdisciplinarity? *Journal of Informetrics*, *13*(3), 904–905. https://doi.org/10.1016/j.joi.2019.03.016

Maggio, L. A., Costello, J. A., Ninkov, A. B., Frank, J. R., & Artino, A. R. (2023). Expanding Interdisciplinarity: A Bibliometric Study of Medical Education Using the Medical Education Journal List-24 (MEJ-24). *Perspectives on Medical Education*, *12*(1), 327–337. https://doi.org/10.5334/pme.984

Marres, N., & De Rijcke, S. (2020). From indicators to indicating interdisciplinarity: A participatory mapping methodology for research communities in-the-making. *Quantitative Science Studies*, *1*(3), 1041–1055. https://doi.org/10.1162/qss_a_00062

McCulloch, W. S., & Pitts, W. (1943). *A logical calculus of the ideas immanent in nervous activity. The bulletin of mathematical biophysics, 5, 115-133.*

Morillo, F., Bordons, M., & Gómez, I. (2001). *An approach to interdisciplinarity through bibliometric indicators*.

Omodei, E., De Domenico, M., & Arenas, A. (2017). Evaluating the impact of interdisciplinary research: A multilayer network approach. *Network Science*, *5*(2), 235–246. https://doi.org/10.1017/nws.2016.15

Porter, A. L., Cohen, A. S., David Roessner, J., & Perreault, M. (2007). Measuring researcher interdisciplinarity. *Scientometrics*, *72*(1), 117–147. https://doi.org/10.1007/s11192-007-1700-5

Porter, A. L., Roessner, D. J., & Heberger, A. E. (2008). How interdisciplinary is a given body of research? *Research Evaluation*, *17*(4), 273–282. https://doi.org/10.3152/095820208X364553

Rafols, I., & Meyer, M. (2010). Diversity and network coherence as indicators of interdisciplinarity: Case studies in bionanoscience. *Scientometrics*, *82*(2), 263–287. https://doi.org/10.1007/s11192-009-0041-y

Ramírez-García, A., González-Molina, A., Gutiérrez-Arenas, M.-P., & Moyano-Pacheco, M. (2022). Interdisciplinarity of scientific production on hate speech and social media: A bibliometric analysis. *Comunicar*, *30*(72), 129–140. https://doi.org/10.3916/C72-2022-10

Rao, C. (1982). Diversity: Its measurement, decomposition, apportionment and analysis. *Sankhy: The Indian Journal of Statistics, Series A, 44(1), 1–22*.

Schmidt, J. C. (2021). *Philosophy of Interdisciplinarity: Studies in Science, Society and Sustainability* (1st ed.). Routledge. https://doi.org/10.4324/9781315387109

Schummer, J. (2004). Multidisciplinarity, interdisciplinarity, and patterns of research collaboration in nanoscience and nanotechnology. *Scientometrics*, *59*(3), 425–465. https://doi.org/10.1023/B:SCIE.0000018542.71314.38

Shen, L., Wang, S., Dai, W., & Zhang, Z. (2019). Detecting the Interdisciplinary Nature and Topic Hotspots of Robotics in Surgery: Social Network Analysis and Bibliometric Study. *J Med Internet Res 2019;21(3):E12625*. https://doi.org/10.2196/12625

Simard, M.-A., Kozlowski, D., Segal, J., Messer, M., Ocay, D. D., Saari, T., Ferland, C. E., & Larivière, V. (2023). Trends in Brain Research: A Bibliometric Analysis. *Canadian Journal of Neurological Sciences*, 1–11. https://doi.org/10.1017/cjn.2023.314

Sjögårde, P., & Ahlgren, P. (2024). Seed-based information retrieval in networks of research publications: Evaluation of direct citations, bibliographic coupling, co-citations, and PubMed- related article score. *Journal of the Association for Information Science and Technology*, *75*(13), 1453–1465. https://doi.org/10.1002/asi.24951

Stirling, A. (2007). A general framework for analysing diversity in science, technology and society. *Journal of The Royal Society Interface*, *4*(15), 707–719. https://doi.org/10.1098/rsif.2007.0213

Sugimoto, C. R., & Larivière, V. (2018). *Measuring research: What everyone needs to know*. Oxford University Press.

Truc, A. (2022). Interdisciplinary influences in behavioral economics: A bibliometric analysis of cross-disciplinary citations. *Journal of Economic Methodology*, *29*(3), 217–251. https://doi.org/10.1080/1350178X.2021.2011374

Vienni Baptista, B., Fletcher, I., Maryl, M., Wciślik, P., Buchner, A., Lyall, C., Spaapen, J., Pohl, C., Klein, J. T., & Schriber, L. (2020). *SHAPE-ID: Shaping Interdisciplinary Practices in Europe Deliverable 2.3: Final Report on Understandings of Interdisciplinary and Transdisciplinary Research and*

*Factors of Success and Failure Project Information*.

Wagner, C. S., Roessner, J. D., Bobb, K., Klein, J. T., Boyack, K. W., Keyton, J., Rafols, I., & Börner, K. (2011). Approaches to understanding and measuring interdisciplinary scientific research (IDR): A review of the literature. *Journal of Informetrics*, *5*(1), 14–26. https://doi.org/10.1016/j.joi.2010.06.004

Wang, Q. (2016). *Studies in the dynamics of science: Exploring emergence, classification, and interdisciplinarity. (Doctoral dissertation, KTH Royal Institute of Technology).* http://urn.kb.se/resolve?urn=urn:nbn:se:kth:diva-184724

Woods, H. B., Rafols, I., & Wilsdon, J. (2024). ***UNDISCIPLINED: How do research funders define transdisciplinary research? (RoRI Working Paper No. 12)***. 16717450 Bytes. https://doi.org/10.6084/M9.FIGSHARE.27088756

Zitt, M., Lelu, A., Cadot, M., & Cabanac, G. (2019). Bibliometric Delineation of Scientific Fields. In W. Glänzel, H. F. Moed, U. Schmoch, & M. Thelwall (Eds.), *Springer Handbook of Science and Technology Indicators* (pp. 25–68). Springer International Publishing. https://doi.org/10.1007/978-3-030-02511-3_2

**Funding**

Diego Kozlowski was financed by SSHRC and MITACS. Malena Méndez Isla was financed by CONICET. Agustín Mauro was financed by CONICET.

**Declarations**

Conflict of interest: The authors have no conflicts of interest to declare to the content of this article.

## Authorship contribution

Conceptualization: MMI, AM, DK

Data curation: MMI, AM

Research: MMI, AM, DK

Methodology: MMI, AM, DK

Project management: DK

Resources: DK

Software: MMI, AM, DK

Supervision: DK

Validation: MMI, AM

Writing - original draft: MMI

Writing - review and editing: MMI, AM, DK

## Data availability

For this research, we used data from the Web of Science (WoS), accessed by the Observatoire des Sciences et des Technologies (https://www.ost.uqam.ca). For legal reasons, data from the Web of Science cannot be made openly available. The code of this project is open available at:

https://github.com/MalenaMendezIsla/retrieval_IDR_neurocomp

If needed, the labeled datasets used in this study can be requested directly from the authors.

## Appendix

### Appendix 1: Keyword Selection Method

We chose absolute frequency in order to select the keyword-ranking. We also tested a relevance-based metric inspired by methods used in topic modeling. Specifically, we calculated keyword relevance as a weighted combination of a keyword's relative frequency in a field and its specificity, defined as the ratio of its relative frequency in the field to its global frequency in WoS. The formula was:

$$\text{relevance}(w|CS) = \lambda \cdot p(w|CS) + (1 - \lambda) \cdot [p(w|CS) / p(w)].$$

We tested three values of λ (0.25, 0.5, and 0.75), and for each case we ranked the top 1,000 keywords accordingly. Although we initially tested ranking by relative frequency within each discipline, this approach led to the inclusion of overly specific terms that only allowed for the inclusion of discipline-specific false positives. Therefore, we opted to use absolute frequency as our selection criterion.

**Appendix 2: Keywords**

**Computer Science Keywords:** Deep learning, machine learning, feature extraction, cloud computing, data mining, Internet of Things, genetic algorithm, computational modeling, feature selection, data models, artificial intelligence, wireless sensor networks, particle swarm optimization, big data, predictive models, fuzzy logic, genetic algorithms, computer vision, blockchain, mathematical model, cryptography, computational complexity, robots, computer architecture, artificial neural networks, support vector machine, approximation algorithms, multi-objective optimization, heuristic algorithms, robotics, IoT, edge computing, virtual reality, cybernetics, servers, distributed systems, analytical models, fuzzy sets, multi-agent systems, adaptation models, software, GPU, parallel computing, model checking, wireless communication, software engineering, image classification, sentiment analysis, Internet, linear programming, principal component analysis, kernel, dynamic programming, hardware, distributed computing, parallel processing, dimensionality reduction, evolutionary algorithms, artificial neural network, mathematical models, access control, throughput, computational geometry, fuzzy control, generative adversarial networks, intrusion detection, recommender systems, wireless sensor network, mobile robots, parallel algorithms, clustering algorithms, solid modeling, wireless networks, collaborative filtering, evolutionary algorithm, human-robot interaction, Java, stability analysis, deep neural network, 5G, bandwidth, prediction algorithms, fog computing, hidden Markov models, network security, evolutionary computation, grid computing, Internet of Things (IoT), finite element method, augmented reality, formal methods, deep neural networks, graph algorithms, biological system modeling, genetic programming, text mining, combinatorial optimization, human-computer interaction, fuzzy clustering, brain modeling.

**Neuroscience Keywords:** Alzheimer's disease, stroke, Parkinson's disease, hippocampus, epilepsy, multiple sclerosis, depression, magnetic resonance imaging, fMRI, dopamine, schizophrenia, cognition, MRI, dementia, brain, pain, apoptosis, neuroprotection, microglia, traumatic brain injury, EEG, neurodegeneration, anxiety, oxidative stress, glutamate, migraine, serotonin, stress, cerebellum, spinal cord, neuroinflammation, attention, spinal cord injury, subarachnoid hemorrhage, immunohistochemistry, striatum, GABA, ischemic stroke, amyotrophic lateral sclerosis, amygdala, prefrontal cortex, neuroimaging, deep brain stimulation, cerebrospinal fluid, mild cognitive impairment, glioma, hypothalamus, basal ganglia, neuropathic pain, blood-brain barrier, neurogenesis, synaptic plasticity, headache, transcranial magnetic stimulation, diffusion tensor imaging, ischemia, seizures, cerebral ischemia, hydrocephalus, chronic pain, retina, glioblastoma, cerebral blood flow, autism, cognitive impairment, aneurysm, electroencephalography, seizure, neuropsychology, neurotoxicity, positron emission tomography, thalamus, functional magnetic resonance imaging, intracerebral hemorrhage, immunocytochemistry, nucleus accumbens, meningioma, central nervous system, neurosurgery, Alzheimer disease, white matter, bipolar disorder, hypoxia, cortex, cocaine, cerebral cortex, neuron, glia, brain injury, motor cortex, neuromodulation, synapse, NMDA receptor, melatonin, addiction, event-related potentials, magnetoencephalography, circadian rhythm, brain tumor.

**Appendix 3: Evaluation of Thresholding Effects on Strategy 3 and Strategy 4**

We tested the effect of applying a threshold for the minimum proportion of citations (for Strategy 3) and references (for Strategy 4) from the opposite field. To evaluate the impact of this thresholding on precision and relative recall, we manually labeled 200 records for each strategy (S3 and S4). Relative recall is defined here as the proportion of true positives out of the total number of positives identified in the full 200-record validation set per strategy. Figure S1 shows the precision, relative recall, F1 score, and number of documents retrieved for each strategy at varying thresholds.

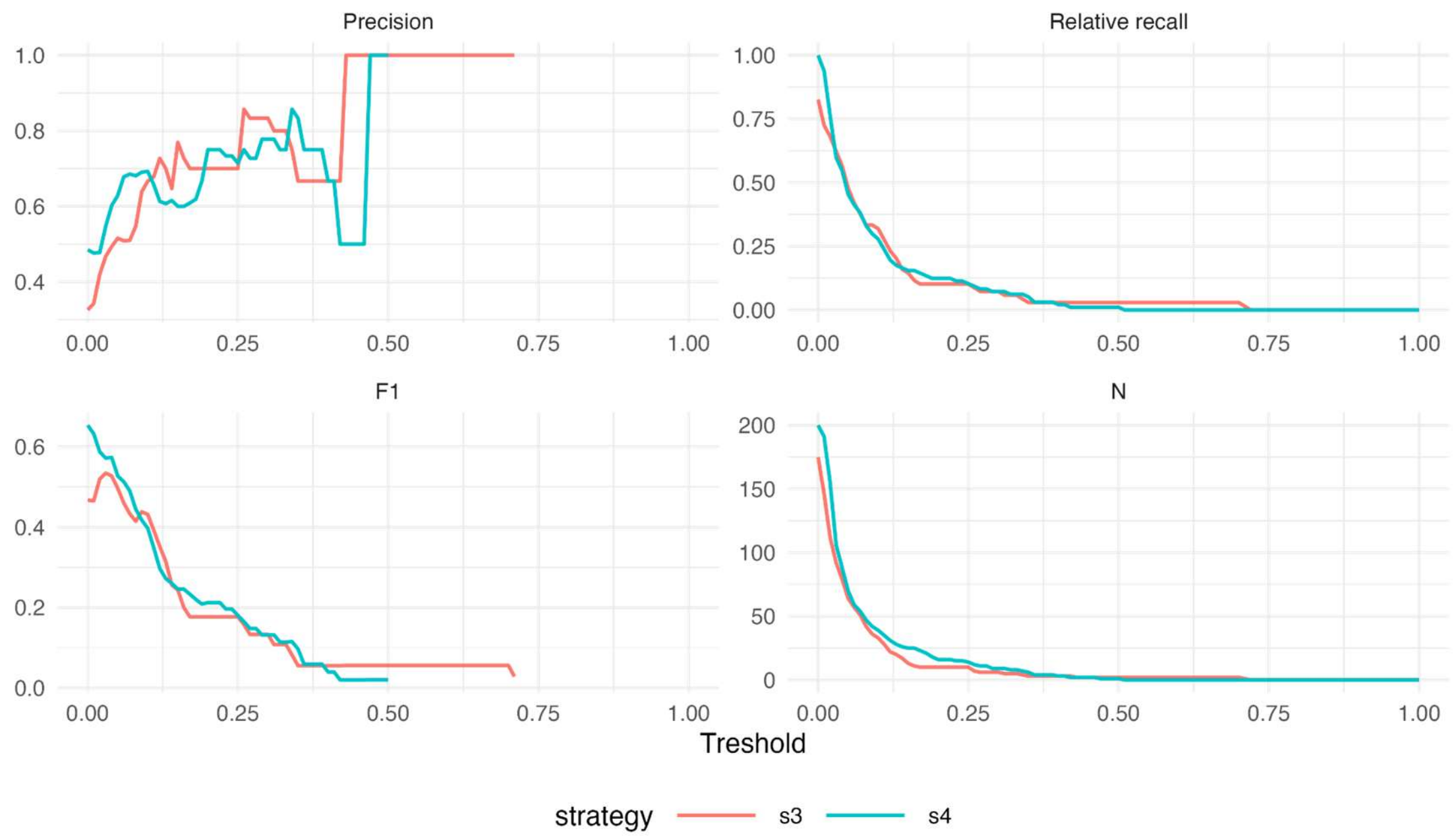

**Figure S1.** Precision and relative recall of Strategies 3 and 4 based on manual validation without thresholding.

This analysis showed that applying thresholds reduces recall more than it improves precision, resulting in a net decrease in F1 score as the threshold increases. Therefore, we decided not to apply any minimum threshold for cross-field citations or references in the final version of the manuscript.

## Appendix 4: Manual Categorization System for Scientific Articles

### 1. Initial Screening:

- **Title and Abstract Review:** Quickly assess if the article potentially integrates neuroscience and computer science.

### 2. In-Depth Evaluation:

#### A. Significant Integration Criterion:

- **Fundamental Integration:** Check if the core objective or findings of the article rely on a substantial combination of data, techniques, tools, concepts, or theories from both neuroscience and computer science.

- **Title/Abstract Reflection:** Confirm if this integration is clearly indicated in the title or abstract.

**B. Article Type Identification:**

- **Neuroscience with Computational Elements:**

  - Determine if computational tools are essential to the article's novelty, contribution, or objective.

  - Ensure the computational tools are not merely routine but play a crucial role (e.g., developing machine learning models, novel software, or brain simulations).

  - Examples: Machine learning used to analyze fMRI or EEG data, decode brain states, and segment stroke lesions, assist in diagnosing, predict the outcome of diseases or interventions. Virtual reality used in the treatment of brain injuries and diseases like Alzheimer's.

- **Computer Science with Elements of Neuroscience:**

  - Assess if the design of algorithms or technologies incorporates neuroscientific research elements.

  - Development of tools for conducting research and gathering data specific for neuroscience.

  - Verify if neuroscientific data or concepts are integral to the design.

  - Examples: Algorithms or robots design based on neuroscience research.

- **Close Interaction:**

  - **Indistinguishable Discipline**: Evaluate if the interaction between neuroscience and computer science is so intertwined that it's challenging to assign the article to a single discipline.

- Examples: Brain-Computer Interface technologies or brain computer simulations that also function as bio-inspired algorithms.

**3. Doubtful Cases:**

- **Collective Decision:** Articles that don't clearly fit the criteria should be flagged and discussed collectively.
- **Focus on Real Interaction:** The decision should be based on identifying a genuine, clear, and valuable interaction between neuroscience and computer science.